\documentclass{article}

\usepackage{amsmath,amssymb,bbm}

\def\kostantop{\not\hspace{-1mm} K}

\begin{document}

\title{Kernel solutions of the Kostant operator on 
		eight-dimensional quotient spaces}

\author{Phongpichit Channuie, Teparksorn Pengpan\footnote{email: teparksorn.p@psu.ac.th} 
~and Witsanu Puttawong
}

\date{}

\maketitle

\centerline{\em Department of Physics, Faculty of Science, 
Prince of Songkla University}
\centerline{\em Hatyai 90112, Thailand}

\vskip 1cm

\begin{abstract}
After introducing the generators and irreducible 
representations of the ${\rm su}(5)$ and ${\rm so}(6)$ Lie algebras in 
terms of the Schwinger's oscillators, the general kernel 
solutions of the Kostant operators on eight-dimensional 
quotient spaces ${\rm su}(5)/ {\rm su}(4)\times {\rm u}(1)$ and ${\rm so}(6)/{\rm so}(4)
\times {\rm so}(2)$ are derived in terms of the diagonal subalgebras 
${\rm su}(4)\times {\rm u}(1)$ and ${\rm so}(4)\times {\rm so}(2)$, respectively.
\end{abstract}

\vskip 1cm

\section{Introduction}

The Dirac operator plays a significant role in quantum field theories.
Its natural generalization with a cubic term arose from Kazama and Suzuki's attempt 
to create a realistic string model \cite{KS}. Their cubic Dirac operator appeared 
in the string model as a supercurrent of a superconformal algebra. Ten years later, 
this type of operator was discovered again by Kostant \cite{Kostant}. He understood 
already that elements in an Euler number multiplet from an equal rank embedding of reductive Lie 
algebras \cite{GKRS} are nothing more than kernel solutions of the cubic Dirac operator. 
It is also an accident that the lowest lines of the Euler number multiplets for the 
4-, 8-, and 16-dimensional coset spaces match with the known supersymmetric 
multiplets \cite{PR}. 

Although the Euler number multiplets are easily derived 
by the GKRS index formula \cite{GKRS}, 
\begin{equation} 
S^+\otimes V_{\Lambda} - S^-\otimes V_{\Lambda}= \sum_{c\in C}
\hbox{sgn}(c)U_{c\bullet \Lambda}, \nonumber
%\label{GKRS}
\end{equation}
they are not helpful for the formulation of any physical theory. In \cite{BRX}, 
Brink, Ramond and Xiong used an algebraic method to determine 
the general kernel solutions or the Euler number multiplets of 
the Kostant operators on the cosets ${\rm SU}(3)/{\rm SU}(2)
\times {\rm U}(1)$ and $F_4/{\rm SO}(9)$. By realizing the 
gamma matrices as dynamical variables satisfying Grassmann 
algebras, the Euler number triplets for ${\rm SU}(3)/{\rm SU}(2)
\times {\rm U}(1)$ and $F_4/{\rm SO}(9)$ were then written 
as chiral superfields. A free action in the light-cone frame for 
both cosets was also formulated.
 
The intention of this paper is to determine the general kernel 
solutions of the Kostant operators on the 8-dimensional quotients 
${\rm su}(5)/{\rm su}(4)\times {\rm u}(1)$ and ${\rm so}(6)
/{\rm so}(4)\times {\rm so}(2)$ by a quantum mechanical method. 
We will briefly present how to construct the generators of 
${\rm su}(5)$ and ${\rm so}(6)$ Lie algebras and their 
irreducible representations (irreps). Only parts that are 
used in constructing the Kostant operators will be mentioned. 
Then, the general kernel solutions will be determined. 
Their extension to the case of a non-compact Lie algebra was 
suggested in 1999 by Ramond from his curiosity to 
know the Euler number multiplets. Some comments about them 
are made in the last section.

\section{Kostant operator of the quotient ${\rm su}(5)/{\rm su}(4)\times {\rm u}(1)$ 
and its kernel solutions} 

\subsection{The Schwinger's oscillator realization of the 
${\rm su}(5)$ Lie algebra}

\begin{figure}[h]

\begin{center}

\begin{picture}(380,160)(0,0)

\footnotesize

%%%%%%%%%%%%%%%%%%%%%%%%%%%%%%%%%%%%%%%%%%%%%%%%%%%%
%  5-dimensional irrep diagram of {\rm su}(5)
%%%%%%%%%%%%%%%%%%%%%%%%%%%%%%%%%%%%%%%%%%%%%%%%%%%%

\multiput(80,130)(0,-20){5}{\circle*{4}}
\multiput(80,125)(0,-20){4}{\vector(0,-1){10}}
\put(85,128){$r_1^\dagger|0>$}
\put(58,119){$-\vec{\alpha}_1$}
\put(85,108){$r_2^\dagger|0>$}
\put(58,99){$-\vec{\alpha}_2$}
\put(85,88){$r_3^\dagger|0>$}
\put(58,79){$-\vec{\alpha}_3$}
\put(85,68){$r_4^\dagger|0>$}
\put(58,59){$-\vec{\alpha}_4$}
\put(85,48){$r_5^\dagger|0>$}

\put(75,5){(a)}

%%%%%%%%%%%%%%%%%%%%%%%%%%%%%%%%%%%%%%%%%%%%%%%%%%%%%%%
%    10-dimensional irrep diagram of {\rm su}(5)
%%%%%%%%%%%%%%%%%%%%%%%%%%%%%%%%%%%%%%%%%%%%%%%%%%%%%%%

\multiput(260,150)(0,-20){2}{\circle*{4}}
\multiput(260,30)(0,20){2}{\circle*{4}}
\multiput(230,110)(0,-20){3}{\circle*{4}}
\multiput(290,110)(0,-20){3}{\circle*{4}}

\multiput(230,105)(0,-20){2}{\vector(0,-1){10}}
\multiput(290,105)(0,-20){2}{\vector(0,-1){10}}
\put(260,145){\vector(0,-1){10}}
\put(252,124){\vector(-2,-1){15}}
\put(268,124){\vector(2,-1){15}}
\put(278,105){\vector(-3,-1){32}}
\put(245,85){\vector(3,-1){32}}
\put(238,65){\vector(2,-1){15}}
\put(282,65){\vector(-2,-1){15}}
\put(260,45){\vector(0,-1){10}}

\put(240,139){$-\vec{\alpha}_2$}
\put(220,120){$-\vec{\alpha}_1$}
\put(282,118){$-\vec{\alpha}_3$}
\put(245,102){$-\vec{\alpha}_1$}
\put(255,82){$-\vec{\alpha}_4$}
\put(222,56){$-\vec{\alpha}_4$}
\put(278,56){$-\vec{\alpha}_2$}
\put(240,38){$-\vec{\alpha}_3$}

\put(195,107){$s_3^\dagger|0>$}
\put(210,98){$-\vec{\alpha}_3$}
\put(195,87){$s_5^\dagger|0>$}
\put(210,78){$-\vec{\alpha}_2$}
\put(195,67){$s_7^\dagger|0>$}
\put(305,107){$s_4^\dagger|0>$}
\put(295,98){$-\vec{\alpha}_4$}
\put(305,87){$s_6^\dagger|0>$}
\put(295,78){$-\vec{\alpha}_1$}
\put(305,67){$s_8^\dagger|0>$}

\put(265,148){$s_1^\dagger|0>$}
\put(265,128){$s_2^\dagger|0>$}
\put(265,47){$s_9^\dagger|0>$}
\put(265,27){$s_{10}^\dagger|0>$}

\put(255,5){(b)}

\end{picture}

\end{center}

\caption{The ${\rm su}(5)$ weight diagrams (a) of a 5-dimensional and (b) of a 
10-dimensional irreps.}
\label{Fig1}
\end{figure}
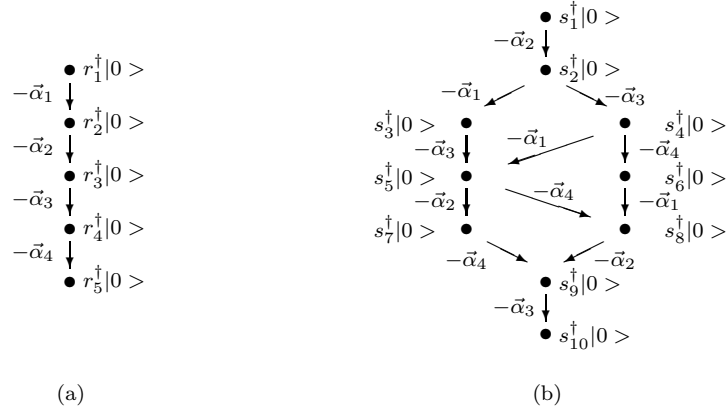

To construct the ${\rm su}(5)$ generators that satisfy Chevalley-Serre 
relations \cite{Fuchs}, we introduce four types of Schwinger's oscillators 
$r_i,~\bar{r}_i,~s_j,~\bar{s}_j$, where $i=1$ to 5 and $j=1$ to 10, including 
their adjoints \cite{Sakurai}. Action of the raising oscillators $r_i^\dagger$ 
and $s_j^\dagger$ on the vacuum state in correspondence to the ${\rm su}(5)$ 
irreps ${\bf 5}$ and ${\bf 10}$ is shown in figure \ref{Fig1} (a) and (b), 
respectively. By reversing all arrows in figure \ref{Fig1} (a) and (b), and replacing 
$r_i^\dagger$ and $s_j^\dagger$ with $\bar{r}_i^\dagger$ and 
$\bar{s}_j^\dagger$, they become the weight diagrams of the $\overline{{\bf 5}}$ and 
$\overline{{\bf 10}}$ irreps. Although the ${\bf 10}$ and $\overline{{\bf 10}}$ 
irreps are not fundamental and can be obtained from anti-symmetrization of ${\bf 5}$ and 
$\overline{{\bf 5}}$ irreps, respectively, it will be seen later that introducing 
the oscillators $s_j$, $\bar{s}_j$ and their adjoints is a convenient 
way in determining the general kernel solutions.             

From the ${\bf 5}$, $\overline{{\bf 5}}$, ${\bf 10}$ and $\overline{{\bf 10}}$ 
weight diagrams, all positive root generators are
\begin{eqnarray}
T_1^+  &=& r_1^\dagger r_2+s_2^\dagger s_3+s_4^\dagger s_5+s_6^\dagger s_8 
+ (r,s\rightarrow \bar{r},\bar{s})^\dagger,
\nonumber\\
T_2^+  &=& r_2^\dagger r_3+s_1^\dagger s_2+s_5^\dagger s_7+s_8^\dagger s_9 
+ (r,s\rightarrow \bar{r},\bar{s})^\dagger,
\nonumber\\
T_3^+  &=& r_1^\dagger r_3-s_1^\dagger s_3+s_4^\dagger s_7+s_6^\dagger s_9 
+ (r,s\rightarrow \bar{r},\bar{s})^\dagger,
\nonumber\\
T_4^+  &=& r_3^\dagger r_4+s_2^\dagger s_4+s_3^\dagger s_5+s_9^\dagger s_{10} 
+ (r,s\rightarrow \bar{r},\bar{s})^\dagger,
\nonumber\\
T_5^+  &=& r_2^\dagger r_4+s_1^\dagger s_4-s_3^\dagger s_7+s_8^\dagger s_{10} 
+ (r,s\rightarrow \bar{r},\bar{s})^\dagger,
\nonumber\\
T_6^+  &=& r_1^\dagger r_4-s_1^\dagger s_5-s_2^\dagger s_7+s_6^\dagger s_{10} 
+ (r,s\rightarrow \bar{r},\bar{s})^\dagger,
\nonumber\\
T_7^+  &=& r_4^\dagger r_5+s_4^\dagger s_6+s_5^\dagger s_8+s_7^\dagger s_9 
+ (r,s\rightarrow \bar{r},\bar{s})^\dagger,
\nonumber\\
T_8^+  &=& r_3^\dagger r_5+s_2^\dagger s_6+s_3^\dagger s_8-s_7^\dagger s_{10} 
+ (r,s\rightarrow \bar{r},\bar{s})^\dagger,
\nonumber\\
T_9^+  &=& r_2^\dagger r_5+s_1^\dagger s_6-s_3^\dagger s_9-s_5^\dagger s_{10} 
+ (r,s\rightarrow \bar{r},\bar{s})^\dagger,
\nonumber\\
T_{10}^+  &=& r_1^\dagger r_5-s_1^\dagger s_8-s_2^\dagger s_9-s_4^\dagger s_{10} 
+ (r,s\rightarrow \bar{r},\bar{s})^\dagger,
\end{eqnarray}
and all negative ones are
\begin{equation}
T_A^- = \left(T_A^+ \right)^\dagger,~~A=1,~2,~3,~ \dots,~10.
\end{equation} 
The Cartan subalgebra generators in the Dynkin basis 
\begin{equation}
\vec{H}\equiv H_1\hat{\omega}_1+H_2\hat{\omega}_2+H_3\hat{\omega}_3
+H_4\hat{\omega}_4=(H_1,H_2,H_3,H_4)
\end{equation}
are obtained from the following commutators:
\begin{eqnarray}
H_1 & \equiv &{[T_1^+ ,T_1^-]} \nonumber\\
	& = & 		N_1^{(r)}+N_2^{(s)}+N_4^{(s)}+N_6^{(s)}
				-N_2^{(r)}-N_3^{(s)}-N_5^{(s)}-N_8^{(s)}
				- (r,s\rightarrow \bar{r},\bar{s}), 
				\nonumber\\
H_2 & \equiv & {[T_2^+ ,T_2^-]} \nonumber\\
	& = &		N_2^{(r)}+N_1^{(s)}+N_5^{(s)}+N_8^{(s)}
				-N_3^{(r)}-N_2^{(s)}-N_7^{(s)}-N_9^{(s)}
				- (r,s\rightarrow \bar{r},\bar{s}), 
				\nonumber\\
H_3 & \equiv & {[T_4^+ ,T_4^-]} \nonumber\\
	& = &		N_3^{(r)}+N_2^{(s)}+N_3^{(s)}+N_9^{(s)}
				-N_4^{(r)}-N_4^{(s)}-N_5^{(s)}-N_{10}^{(s)}
				- (r,s\rightarrow \bar{r},\bar{s}), 
				\nonumber\\
H_4 & \equiv & {[T_7^+ ,T_7^-]} \nonumber\\
	& = &		N_4^{(r)}+N_4^{(s)}+N_5^{(s)}+N_7^{(s)}
				-N_5^{(r)}-N_6^{(s)}-N_8^{(s)}-N_9^{(s)}
				- (r,s\rightarrow \bar{r},\bar{s}), 			
\end{eqnarray}
where $N_i^{r,s,\bar{r},\bar{s}}$ are the number operators. 
They are related to the Cartan generators in the orthonormal basis 
\begin{equation}
\vec{h}\equiv h_1\hat{e}_1+h_2\hat{e}_2+h_3\hat{e}_3
+h_4\hat{e}_4+h_5\hat{e}_5=[h_1,h_2,h_3,h_4,h_5]
\end{equation}
as follows:
\begin{equation}
H_1=h_1-h_2,~H_2=h_2-h_3,~H_3=h_3-h_4,~H_4=h_4-h_5. 
\end{equation}

An ${\rm su}(5)$ irrep represented by its highest weight $\Lambda$ in 
its vector space $V_\Lambda$ can be generated by 
action of the raising oscillators, $r_1^\dagger,~\bar{r}_5^\dagger,
~s_1^\dagger,~\bar{s}_{10}^\dagger$, on the vacuum state  
\begin{equation}
\Lambda = (r_1^\dagger )^{a_1}(s_1^\dagger )^{a_2}
(\bar{s}_{10}^\dagger )^{a_3}(\bar{r}_5^\dagger )^{a_4}|0>, 
\end{equation}
where $a_{1,2,3,4}$ are non-negative integers, called the Dynkin labels. 
The action of the Cartan generators in the Dynkin basis on the 
highest weight gives their eigenvalues as follow:
\begin{equation}
\vec{H}\Lambda=(a_1,a_2,a_3,a_4)\Lambda,
\end{equation}
and in the orthonormal basis as follow:
\begin{equation}
\vec{h}\Lambda=[b_1,b_2,b_3,b_4,b_5]\Lambda,
\end{equation}  
 where
\begin{eqnarray}
b_1 &=& \frac 15(4a_1+3a_2+2a_3+a_4), \nonumber\\
b_2 &=& \frac 15(-a_1+3a_2+2a_3+a_4), \nonumber\\
b_3 &=& \frac 15(-a_1-2a_2+2a_3+a_4), \nonumber\\
b_4 &=& \frac 15(-a_1-2a_2-3a_3+a_4), \nonumber\\
b_5 &=& \frac 15(-a_1-2a_2-3a_3-4a_4).
\end{eqnarray}
Note that $b_1+b_2+b_3+b_4+b_5=0$ is due to the basis constraint.

Inside the ${\rm su}(5)$ generators, the generators $T_{1,2,\dots,6}^\pm $ 
and $H_{1,2,3}$ form the ${\rm su}(4)$  Lie subalgebra and the generator 
$h_5$ is the generator of ${\rm u}(1)$ subalgebra. The other generators 
$T_{7,8,9,10}^\pm$ lie outside the subalgebra ${\rm su}(4)\times {\rm u}(1)$ 
and are used to construct the Kostant operator of the quotient 
${\rm su}(5)/{\rm su}(4)\times {\rm u}(1)$.      

\subsection{Kernel solutions of the Kostant operator}

To construct the Kostant operator on the 8-dimensional quotient space, 
the following $16\times 16$ gamma matrices are needed: 
\begin{eqnarray}
%\label{002}
\Gamma_1 &=&\sigma_1\otimes\sigma_1\otimes\sigma_1\otimes\sigma_1,
\quad
\Gamma_5 =\sigma_1\otimes\sigma_1\otimes\sigma_3\otimes\mathbbm{1},
\nonumber\\
\Gamma_2 &=&\sigma_1\otimes\sigma_1\otimes\sigma_1\otimes\sigma_2,
\quad
\Gamma_6 =\sigma_1\otimes\sigma_2\otimes\mathbbm{1}\otimes\mathbbm{1},
\nonumber\\
\Gamma_3 &=&\sigma_1\otimes\sigma_1\otimes\sigma_1\otimes\sigma_3,
\quad
\Gamma_7 =\sigma_1\otimes\sigma_3\otimes\mathbbm{1}\otimes\mathbbm{1},
\nonumber\\
\Gamma_4 &=&\sigma_1\otimes\sigma_1\otimes\sigma_2\otimes\mathbbm{1},
\quad~
\Gamma_8 =\sigma_2\otimes\mathbbm{1}\otimes\mathbbm{1}\otimes\mathbbm{1},
\end{eqnarray}
where $\sigma_{1,2,3}$ are the Pauli matrices and 
$\mathbbm{1}$ is a $2\times 2$ identity matrix. 
These gamma matrices satisfy Clifford algebra
\begin{equation}
\{\Gamma_a,\Gamma_b\}=2\delta_{a,b}
\left(\mathbbm{1}\otimes\mathbbm{1}\otimes\mathbbm{1}\otimes\mathbbm{1}\right).
\end{equation}
To associate with the generators of the quotient 
${\rm su}(5)/{\rm su}(4)\times {\rm u}(1)$, 
the gamma matrices are complexified as follows:
\begin{eqnarray}
%\label{002}
\gamma_7^{\pm} &=& \frac 12(\Gamma_1\pm i\Gamma_2)
=\sigma_1\otimes\sigma_1\otimes\sigma_1\otimes\sigma^\pm,
\nonumber\\
\gamma_8^{\pm} &=& \frac 12(\Gamma_3\pm i\Gamma_4)
=\sigma_1\otimes\sigma_1\otimes
{[\sigma^+\otimes\frac 12(\sigma_3\pm\mathbbm{1})
+\sigma^-\otimes\frac 12(\sigma_3\mp\mathbbm{1})]},
\nonumber\\
\gamma_9^{\pm} &=& \frac 12(\Gamma_5\pm i\Gamma_6)
=\sigma_1\otimes{[\sigma^+\otimes
\frac 12(\sigma_3\pm\mathbbm{1})+\sigma^-\otimes
\frac 12(\sigma_3\mp\mathbbm{1})]}\otimes\mathbbm{1},
\nonumber\\
\gamma_{10}^{\pm} &=& \frac 12(\Gamma_7\pm i\Gamma_8)
={[\sigma^+\otimes
\frac 12(\sigma_3\pm\mathbbm{1})+\sigma^-\otimes
\frac 12(\sigma_3\mp\mathbbm{1})]}\otimes\mathbbm{1}\otimes\mathbbm{1}.
\end{eqnarray}
Under these complexification, the positive spinor states of ${\rm so}(8)$ 
are $|+\pm\pm\pm\!\!>$ and the negative ones $|-\pm\pm\pm\!\!>$.

From the commutators of the generators of the quotient,
\begin{eqnarray}
{[T_7^+ ,T_7^-]}&=& h_4-h_5,~~~~
{[T_8^+ ,T_8^-]} =  h_3-h_5,\nonumber\\
{[T_9^+ ,T_9^-]}&=&h_2-h_5,~~~~
{[T_{10}^+ ,T_{10}^-]} = h_1-h_5,
\label{structure_constant1}
\end{eqnarray}
the generators $T_{a=7,8,9,10}^\pm$ are not generated. 
The structure constants of which transformations 
are zero. Hence, there are no cubic terms, which are composed of a product 
of three gamma matrices associated with the structure constants.  
The Kostant operator of the quotient ${\rm su}(5)/{\rm su}(4)\times {\rm u}(1)$ is just
\begin{equation}
\kostantop = \sum_{a=7}^{10} (\gamma_a^+ T_a^- + \gamma_a^- T_a^+).
\end{equation}
This Kostant operator acts on a tensor-product space of 
the ${\rm so}(8)$ spinor representations and the ${\rm su}(5)$ irrep   
\begin{equation}
\psi_\Lambda^{\pm}\equiv |\pm\pm\pm\pm\!\!>\! V_\Lambda, 
\end{equation}
and there exist kernel solutions such that
\begin{equation}
\label{Kostant1}
\kostantop\psi_{\lambda_i}^{\pm} =0,
\end{equation}
where $\lambda_i$ is a weight in the vector space $V_\Lambda$. 
Equation (\ref{Kostant1}) can be decomposed into sixteen, 
independent equations as follows:
\begin{eqnarray}
(T_7^+  + T_8^+  + T_9^+  + T_{10}^+ )
\psi_{\lambda_1}^{+}&=0,
\nonumber\\
(T_7^- - T_8^- + T_9^+  + T_{10}^+ )
\psi_{\lambda_2}^{+}&=0,
\nonumber\\
(T_7^+  + T_8^- - T_9^- + T_{10}^+ )
\psi_{\lambda_3}^{+}&=0, 
\nonumber\\
(T_7^- - T_8^+  - T_9^- + T_{10}^+ )
\psi_{\lambda_4}^{+}&=0,
\nonumber\\
(T_7^+  + T_8^+  + T_9^- - T_{10}^-)
\psi_{\lambda_5}^{+}&=0, 
\nonumber\\
(T_7^- - T_8^- + T_9^- - T_{10}^-)
\psi_{\lambda_6}^{+}&=0, 
\nonumber\\
(T_7^+  + T_8^- - T_9^+  - T_{10}^-)
\psi_{\lambda_7}^{+}&=0,
\nonumber\\
(T_7^- - T_8^+  - T_9^+  - T_{10}^-)
\psi_{\lambda_8}^{+}&=0,
\nonumber\\
(T_7^+  + T_8^+  + T_9^+  + T_{10}^-)
\psi_{\lambda_1^\prime}^{-}&=0, 
\nonumber\\
(T_7^- - T_8^- + T_9^+  + T_{10}^-)
\psi_{\lambda_2^\prime}^{-}&=0, 
\nonumber\\
(T_7^+  + T_8^- - T_9^- + T_{10}^-)
\psi_{\lambda_3^\prime}^{-}&=0, 
\nonumber\\
(T_7^- - T_8^+  - T_9^- + T_{10}^-)
\psi_{\lambda_4^\prime}^{-}&=0, 
\nonumber\\
(T_7^+  + T_8^+  + T_9^- - T_{10}^+ )
\psi_{\lambda_5^\prime}^{-}&=0, 
\nonumber\\
(T_7^- - T_8^- + T_9^- - T_{10}^+ )
\psi_{\lambda_6^\prime}^{-}&=0, 
\nonumber\\
(T_7^+  + T_8^- - T_9^+  - T_{10}^+ )
\psi_{\lambda_7^\prime}^{-}&=0, 
\nonumber\\
(T_7^- - T_8^+  - T_9^+  - T_{10}^+ )
\psi_{\lambda_8^\prime}^{-}&=0.   
\end{eqnarray}
One of the possible kernel solutions in the positive spinor space is 
as follows:
\begin{eqnarray}
%\label{002}
\psi_{\lambda_1}^{+}&=&|++++> 
(r_1^\dagger )^{a_1}(s_1^\dagger )^{a_2}(\bar{s}_{10}^\dagger )^{a_3}
(\bar{r}_5^\dagger )^{a_4}|0>,
\nonumber\\
\psi_{\lambda_2}^{+}&=&|+++-> 
(r_1^\dagger )^{a_1}(s_1^\dagger )^{a_2}(\bar{s}_{7}^\dagger )^{a_3}
(\bar{r}_4^\dagger )^{a_4}|0>,
\nonumber\\
\psi_{\lambda_3}^{+}&=&|++-+> 
(r_1^\dagger )^{a_1}(s_4^\dagger )^{a_2}(\bar{s}_3^\dagger )^{a_3}
(\bar{r}_3^\dagger )^{a_4}|0>, 
\nonumber\\
\psi_{\lambda_4}^{+}&=&|++--> 
(r_1^\dagger )^{a_1}(s_2^\dagger )^{a_2}(\bar{s}_5^\dagger )^{a_3}
(\bar{r}_4^\dagger )^{a_4}|0>,
\nonumber\\
\psi_{\lambda_5}^{+}&=&|+-++> 
(r_4^\dagger )^{a_1}(s_7^\dagger )^{a_2}(\bar{s}_1^\dagger )^{a_3}
(\bar{r}_1^\dagger )^{a_4}|0>, 
\nonumber\\
\psi_{\lambda_6}^{+}&=&|+-+-> 
(r_5^\dagger )^{a_1}(s_6^\dagger )^{a_2}(\bar{s}_7^\dagger )^{a_3}
(\bar{r}_4^\dagger )^{a_4}|0>,
\nonumber\\
\psi_{\lambda_7}^{+}&=&|+--+> 
(r_2^\dagger )^{a_1}(s_5^\dagger )^{a_2}(\bar{s}_2^\dagger )^{a_3}
(\bar{r}_1^\dagger )^{a_4}|0>,
\nonumber\\
\psi_{\lambda_8}^{+}&=&|+---> 
(r_3^\dagger )^{a_1}(s_3^\dagger )^{a_2}(\bar{s}_4^\dagger )^{a_3}
(\bar{r}_4^\dagger )^{a_4}|0>, 
\end{eqnarray}
and in the negative spinor space as follows:
\begin{eqnarray}
%\label{002}
\psi_{\lambda_1^\prime}^{-}&=&|-+++> 
(r_4^\dagger )^{a_1}(s_7^\dagger )^{a_2}(\bar{s}_6^\dagger )^{a_3}
(\bar{r}_1^\dagger )^{a_4}|0>, 
\nonumber\\
\psi_{\lambda_2^\prime}^{-}&=&|-++-> 
(r_2^\dagger )^{a_1}(s_8^\dagger )^{a_2}(\bar{s}_7^\dagger )^{a_3}
(\bar{r}_4^\dagger )^{a_4}|0>,
\nonumber\\
\psi_{\lambda_3^\prime}^{-}&=&|-+-+> 
(r_4^\dagger )^{a_1}(s_{10}^\dagger )^{a_2}(\bar{s}_1^\dagger )^{a_3}
(\bar{r}_1^\dagger )^{a_4}|0>, 
\nonumber\\
\psi_{\lambda_4^\prime}^{-}&=&|-+--> 
(r_3^\dagger )^{a_1}(s_9^\dagger )^{a_2}(\bar{s}_5^\dagger )^{a_3}
(\bar{r}_4^\dagger )^{a_4}|0>,
\nonumber\\
\psi_{\lambda_5^\prime}^{-}&=&|--++> 
(r_4^\dagger )^{a_1}(s_7^\dagger )^{a_2}(\bar{s}_8^\dagger )^{a_3}
(\bar{r}_2^\dagger )^{a_4}|0>, 
\nonumber\\
\psi_{\lambda_6^\prime}^{-}&=&|--+-> 
(r_1^\dagger )^{a_1}(s_6^\dagger )^{a_2}(\bar{s}_7^\dagger )^{a_3}
(\bar{r}_4^\dagger )^{a_4}|0>,
\nonumber\\
\psi_{\lambda_7^\prime}^{-}&=&|---+> 
(r_4^\dagger )^{a_1}(s_5^\dagger )^{a_2}(\bar{s}_9^\dagger )^{a_3}
(\bar{r}_3^\dagger )^{a_4}|0>,
\nonumber\\
\psi_{\lambda_8^\prime}^{-}&=&|----> 
(r_1^\dagger )^{a_1}(s_1^\dagger )^{a_2}(\bar{s}_{10}^\dagger )^{a_3}
(\bar{r}_4^\dagger )^{a_4}|0>.
\end{eqnarray}

To get the kernel solutions in terms of ${\rm su}(4)\times {\rm u}(1)$, 
it needs to act on them by the Cartan subalgebra 
generators, which in the Dynkin basis are 
\begin{eqnarray}
\label{diagonal_eq1}
D_1&=& h_1-h_2 + \frac 12\left(f_{+-1}^{10}[\gamma_{10}^+ ,\gamma_{10}^-] 
              - f_{+-2}^9[\gamma_9^+ ,\gamma_9^-]\right)
\nonumber\\
		&=& H_1 + \frac 12\left(
	   			\sigma_3\otimes\sigma_3\otimes\mathbbm{1}\otimes\mathbbm{1} 
	   			- \mathbbm{1}\otimes\sigma_3\otimes\sigma_3\otimes\mathbbm{1}\right),
\nonumber\\
D_2&=& h_2-h_3 + \frac 12\left(
 		 	f_{+-2}^9[\gamma_9^+ ,\gamma_9^-] 
         	- f_{+-3}^8[\gamma_8^+ ,\gamma_8^-]\right)
\nonumber\\
		&=& H_2 + \frac 12\left(
	   		\mathbbm{1}\otimes\sigma_3\otimes\sigma_3\otimes\mathbbm{1}
	   		- \mathbbm{1}\otimes\mathbbm{1}\otimes\sigma_3\otimes\sigma_3\right),
\nonumber\\
D_3 &=& h_3-h_4+ \frac 12\left(f_{+-3}^8[\gamma_8^+ ,\gamma_8^-] 
       - f_{+-4}^7[\gamma_7^+ ,\gamma_7^-]\right)
\nonumber\\
	&=& H_3 + \frac 12\left( 
	     \mathbbm{1}\otimes\mathbbm{1}\otimes\sigma_3\otimes\sigma_3
	   - \mathbbm{1}\otimes\mathbbm{1}\otimes\mathbbm{1}\otimes\sigma_3\right),
\nonumber\\
D_4 &=& \frac 12 h_5 
		+ \frac 14\left(f_{+-5}^7[\gamma_7^+ ,\gamma_7^-] 
		+ f_{+-5}^8[\gamma_8^+ ,\gamma_8^-]
		+f_{+-5}^9[\gamma_9^+ ,\gamma_9^-] 
		+ f_{+-5}^{10}[\gamma_{10}^+ ,\gamma_{10}^-]\right)
\nonumber\\
	&=& \frac 12 h_5 - \frac 14( \mathbbm{1}\otimes\mathbbm{1}\otimes\mathbbm{1}\otimes\sigma_3
		   +\mathbbm{1}\otimes\mathbbm{1}\otimes\sigma_3\otimes\sigma_3
		   + \mathbbm{1}\otimes\sigma_3\otimes\sigma_3\otimes\mathbbm{1} 
		   \nonumber\\
	&&~~~~~
	   + \sigma_3\otimes\sigma_3\otimes\mathbbm{1}\otimes\mathbbm{1}).
\end{eqnarray}
The structure constants in (\ref{diagonal_eq1}) are read directly from 
(\ref{structure_constant1}). The generators $D_1,~D_2$ and $D_3$ 
are the Cartan generators of ${\rm su}(4)$ and the generator 
$D_4$ is the Cartan generator of ${\rm u}(1)$. When the Cartan generators 
act on the kernel solutions, they yield
\begin{eqnarray}
\label{euler_multiplet_su5}
(D_1,D_2,D_3;D_4)\psi_{\lambda_1}^{+} 
 & = & (a_1,a_2,a_3;(b_5-2)/2)\psi_{\lambda_1}^{+},
\nonumber\\
(D_1,D_2,D_3;D_4)\psi_{\lambda_2}^{+}
 & = & (a_1,a_2+a_3+1,a_4;b_3/2)\psi_{\lambda_2}^{+},
\nonumber\\
(D_1,D_2,D_3;D_4)\psi_{\lambda_3}^{+}
 & = & (a_1+a_2+a_3+1,a_4,-a_2-a_3-a_4-1; b_3/2)\psi_{\lambda_3}^{+},
\nonumber\\
(D_1,D_2,D_3;D_4)\psi_{\lambda_4}^{+}
 & = & (a_1+a_2+a_3+1,-a_2-a_3-1,a_2+a_3+a_4+1;b_3/2)\psi_{\lambda_4}^{+},
\nonumber\\
(D_1,D_2,D_3;D_4)\psi_{\lambda_5}^{+}
 & = & (-a_4,-a_2-a_3-1,-a_1;b_3/2)\psi_{\lambda_5}^{+},
\nonumber\\
(D_1,D_2,D_3;D_4)\psi_{\lambda_6}^{+}
 & = & (a_2,a_3,a_4;(b_1+2)/2)\psi_{\lambda_6}^{+},
\nonumber\\
(D_1,D_2,D_3;D_4)\psi_{\lambda_7}^{+}
 & = & (-a_1-a_2-a_3-a_4-1,a_1+a_2+a_3+1,-a_2-a_3-1;b_3/2)\psi_{\lambda_7}^{+},
\nonumber\\
(D_1,D_2,D_3;D_4)\psi_{\lambda_8}^{+}
 & = & (-a_2-a_3-1,-a_1,a_1+a_2+a_3+a_4+1;b_3/2)\psi_{\lambda_8}^{+},
\nonumber\\
(D_1,D_2,D_3;D_4)\psi_{\lambda_1^\prime}^{-}
 & = & (-a_3-a_4-1,-a_2,-a_1;(b_4-1)/2 )\psi_{\lambda_1^\prime}^{-},
\nonumber\\
(D_1,D_2,D_3;D_4)\psi_{\lambda_2^\prime}^{-}
 & = & (-a_1-a_2-1,a_1+a_2+a_3+1,a_4;(b_2+1)/2 )\psi_{\lambda_2^\prime}^{-},
\nonumber\\
(D_1,D_2,D_3;D_4)\psi_{\lambda_3^\prime}^{-}
 & = & (-a_4,-a_3,-a_1-a_2-1;(b_2+1)/2)\psi_{\lambda_3^\prime}^{-},
\nonumber\\
(D_1,D_2,D_3;D_4)\psi_{\lambda_4^\prime}^{-}
 & = & (a_3,-a_1-a_2-a_3-1,a_1+a_2+a_3+a_4+1;(b_2+1)/2 )\psi_{\lambda_4^\prime}^{-},
\nonumber\\
(D_1,D_2,D_3;D_4)\psi_{\lambda_5^\prime}^{-}
 & = & (a_3+a_4+1,-a_2-a_3-a_4-1,-a_1;(b_4-1)/2 )\psi_{\lambda_5^\prime}^{-},
\nonumber\\
(D_1,D_2,D_3;D_4)\psi_{\lambda_6^\prime}^{-}
 & = & (a_1+a_2+1,a_3,a_4;(b_2+1)/2)\psi_{\lambda_6^\prime}^{-},
\nonumber\\
(D_1,D_2,D_3;D_4)\psi_{\lambda_7^\prime}^{-}
 & = & (-a_2,a_2+a_3+a_4+1,-a_1-a_2-a_3-a_4-1;(b_4-1)/2)\psi_{\lambda_7^\prime}^{-},
\nonumber\\
(D_1,D_2,D_3;D_4)\psi_{\lambda_8^\prime}^{-}
 & = & (a_1,a_2,a_3+a_4+1;(b_4-1)/2)\psi_{\lambda_8^\prime}^{-}.
\end{eqnarray}
In case $a_1=a_2=a_3=a_4=0$, the kernel solutions 
(\ref{euler_multiplet_su5}) can be grouped in terms of ${\rm su}(4)$ dimensions 
as follows:
\begin{equation}
{\bf 1}_{-1}\equiv \psi_{\lambda_1}^{+} \sim (0,0,0)_{-1},
\quad
{\bf 6}_0\equiv \left\{ 
\begin{array}{l}
\psi_{\lambda_2}^{+} \sim (~0,~1,~0)_0 \\
\psi_{\lambda_4}^{+} \sim (~1,-1,~1)_0 \\
\psi_{\lambda_8}^{+} \sim (-1,~0,~1)_0 \\
\psi_{\lambda_3}^{+} \sim (~1,~0,-1)_0 \\
\psi_{\lambda_7}^{+} \sim (-1,~1,-1)_0 \\
\psi_{\lambda_5}^{+} \sim (~0,-1,~0)_0 \\ 
\end{array}
\right.,
\quad
{\bf 1}_{1}\equiv \psi_{\lambda_6}^{+} \sim (0,0,0)_{1},
\nonumber
\end{equation}
\begin{equation}
{\bf 4}_{-1/2}\equiv 
\left\{ 
\begin{array}{l}
\psi_{\lambda_1^\prime}^{-} \sim (-1,~0,~0)_{-1/2} \\
\psi_{\lambda_5^\prime}^{-} \sim (~1,-1,~0)_{-1/2} \\
\psi_{\lambda_7^\prime}^{-} \sim (~0,~1,-1)_{-1/2} \\
\psi_{\lambda_8^\prime}^{-} \sim (~0,~0,~1)_{-1/2} 
\end{array}
\right.,
\quad\quad
{\bf 4}_{1/2}\equiv 
\left\{ 
\begin{array}{l}
\psi_{\lambda_6^\prime}^{-} \sim (~1,~0,~0)_{1/2} \\
\psi_{\lambda_2^\prime}^{-} \sim (-1,~1,~0)_{1/2} \\
\psi_{\lambda_4^\prime}^{-} \sim (~0,-1,~1)_{1/2} \\
\psi_{\lambda_3^\prime}^{-} \sim (~0,~0,-1)_{1/2} 
\end{array}
\right..
\nonumber
\end{equation}
Since the Dynkin labels $a_{1,2,3,4}$ are non-negative, the direct sum of 
the ${\rm su}(4)$ highest weights 
\begin{equation} 
\psi_{\lambda_1}^{+}
\oplus \psi_{\lambda_2}^{+}
\oplus \psi_{\lambda_6}^{+}
\oplus \psi_{\lambda_8^\prime}^{-}
\oplus \psi_{\lambda_6^\prime}^{-} 
\end{equation}
forms the Euler number multiplet. In terms of its Dynkin labels, it read:
\begin{eqnarray} 
&
(a_1,a_2,a_3)_{(b_5-2)/2}
\oplus (a_1,a_2+a_3+1,a_4)_{b_3/2}
\oplus (a_2,a_3,a_4)_{(b_1+2)/2}
\nonumber\\
&
\oplus (a_1,a_2,a_3+a_4+1)_{(b_4-1)/2}
\oplus (a_1+a_2+1,a_3,a_4)_{(b_2+1)/2}. 
\end{eqnarray}

\section{Kostant operator of the quotient ${\rm so}(6)/{\rm so}(4)\times {\rm so}(2)$ 
and its kernel solutions} 

\subsection{The Schwinger's oscillator realization of the ${\rm so}(6)$ Lie algebra}

\begin{figure}[h]

\begin{center}

\begin{picture}(380,120)(0,0)

\footnotesize

%%%%%%%%%%%%%%%%%%%%%%%%%%%%%%%%%%%%%%%%%%%%%%%%%%%%%%%
%    6-dimensional irrep diagram of {\rm so}(6)
%%%%%%%%%%%%%%%%%%%%%%%%%%%%%%%%%%%%%%%%%%%%%%%%%%%%%%%

\put(80,110){\circle*{4}}
\put(80,90){\circle*{4}}
\put(50,70){\circle*{4}}
\put(110,70){\circle*{4}}
\put(80,50){\circle*{4}}
\put(80,30){\circle*{4}}

\put(80,105){\vector(0,-1){10}}
\put(72,84){\vector(-2,-1){15}}
\put(88,84){\vector(2,-1){15}}
\put(58,64){\vector(2,-1){15}}
\put(102,64){\vector(-2,-1){15}}
\put(80,45){\vector(0,-1){10}}

\put(60,99){$-\vec{\alpha}_1$}
\put(40,80){$-\vec{\alpha}_2$}
\put(102,78){$-\vec{\alpha}_3$}
\put(40,55){$-\vec{\alpha}_3$}
\put(102,56){$-\vec{\alpha}_2$}
\put(60,39){$-\vec{\alpha}_1$}

\put(85,108){$r_1^\dagger|0>$}
\put(85,88){$r_2^\dagger|0>$}
\put(20,68){$r_3^\dagger|0>$}
\put(120,68){$\bar{r}_3^\dagger|0>$}
\put(85,46){$\bar{r}_2^\dagger|0>$}
\put(85,28){$\bar{r}_1^\dagger|0>$}

\put(75,5){(a)}
%%%%%%%%%%%%%%%%%%%%%%%%%%%%%%%%%%%%%%%%%%%%%%%%%%%%
%  4-dimensional co-spinor diagram of {\rm so}(6)
%%%%%%%%%%%%%%%%%%%%%%%%%%%%%%%%%%%%%%%%%%%%%%%%%%%%

\multiput(220,100)(0,-20){4}{\circle*{4}}
\multiput(220,95)(0,-20){3}{\vector(0,-1){10}}

\put(200,89){$-\vec{\alpha}_2$}
\put(200,69){$-\vec{\alpha}_1$}
\put(200,49){$-\vec{\alpha}_3$}

\put(225,98){$s_1^\dagger|0>$}
\put(225,78){$s_2^\dagger|0>$}
\put(225,58){$s_3^\dagger|0>$}
\put(225,38){$s_4^\dagger|0>$}

\put(215,5){(b)}

%%%%%%%%%%%%%%%%%%%%%%%%%%%%%%%%%%%%%%%%%%%%%%%%%%%%
%  4-dimensional spinor diagram of {\rm so}(6)
%%%%%%%%%%%%%%%%%%%%%%%%%%%%%%%%%%%%%%%%%%%%%%%%%%%%

\multiput(340,100)(0,-20){4}{\circle*{4}}
\multiput(340,95)(0,-20){3}{\vector(0,-1){10}}

\put(320,89){$-\vec{\alpha}_3$}
\put(320,69){$-\vec{\alpha}_1$}
\put(320,49){$-\vec{\alpha}_2$}

\put(345,98){$\bar{s}_4^\dagger|0>$}
\put(345,78){$\bar{s}_3^\dagger|0>$}
\put(345,58){$\bar{s}_2^\dagger|0>$}
\put(345,38){$\bar{s}_1^\dagger|0>$}

\put(335,5){(c)}

\end{picture}

\end{center}

\caption{The ${\rm so}(6)$ weight diagrams (a) of a 6-dimensional vector 
(b) of a 4-dimensional co-spinor and (c) of a 4-dimensional spinor 
representations.}
\label{Fig2}
\end{figure}
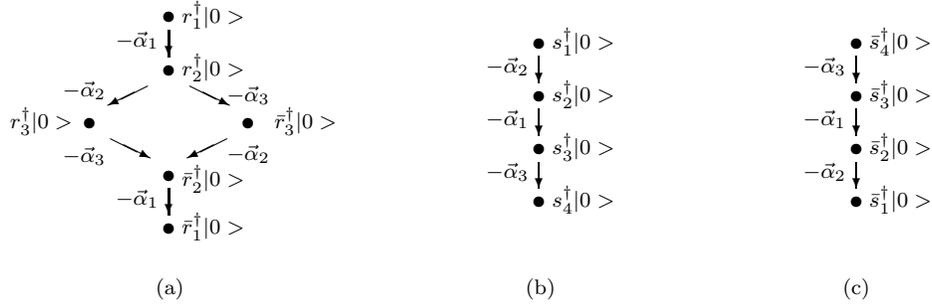

To construct the generators for ${\rm so}(6)$, we introduce four types of Schwinger's 
oscillators $r_i,~\bar{r}_i,~s_j,~\bar{s}_j$, where $i=1,2,3$ and $j=1,2,3,4$, 
including their adjoints. Action of the raising oscillators $r_i^\dagger,
~\bar{r}_i^\dagger,~s_i^\dagger$ and $\bar{s}_j^\dagger$ on the vacuum state 
in correspondence to the ${\rm so}(6)$ irreps ${\bf 6}$, ${\bf 4}_c$ and 
${\bf 4}_s$ is shown in figure \ref{Fig2} (a), (b) and (c), respectively. 
Although the ${\bf 6}$ irrep is not fundamental and can be obtained from 
an anti-symmetric product of two copies of either the ${\bf 4}_c$ or 
${\bf 4}_s$ irrep, it will be seen later that introducing the oscillators 
$r_j$ and $\bar{r}_j$ is an easy way to determine the kernel 
solutions of the Kostant operator.        

From the weight diagrams of ${\rm so}(6)$, all positive root generators are
\begin{eqnarray}
T_1^+  &=& r_1^\dagger r_2+s_2^\dagger s_3
+ (r,s\rightarrow \bar{r},\bar{s})^\dagger,
\nonumber\\
T_2^+  &=& r_2^\dagger r_3+s_1^\dagger s_2
+ (r,s\rightarrow \bar{r},\bar{s})^\dagger,
\nonumber\\
T_3^+  &=& r_2^\dagger \bar{r}_3+s_3^\dagger s_4 
+ (r,\bar{r},s\rightarrow \bar{r},r,\bar{s})^\dagger,
\nonumber\\
T_4^+  &=& r_1^\dagger r_3-s_1^\dagger s_3
- (r,s\rightarrow \bar{r},\bar{s})^\dagger,
\nonumber\\
T_5^+  &=& r_1^\dagger \bar{r}_3+s_2^\dagger s_4 
- (r,\bar{r},s\rightarrow \bar{r},r,\bar{s})^\dagger,
\nonumber\\
T_6^+  &=& -r_1^\dagger \bar{r}_2+s_1^\dagger s_4 
+ (r,\bar{r},s\rightarrow \bar{r},r,\bar{s})^\dagger,
\end{eqnarray}
and all negative root generators are
\begin{equation}
T_A^- = \left(T_A^+ \right)^\dagger,~~A=1,~2,~3,~ \dots,~6.
\end{equation} 
The Cartan subalgebra generators in the Dynkin basis, 
\begin{equation}
\vec{H}\equiv H_1\hat{\omega}_1+H_2\hat{\omega}_2+H_3\hat{\omega}_3
=(H_1,H_2,H_3),
\end{equation}
are obtained from the following commutators:
\begin{eqnarray}
H_1 &\equiv& {[T_1^+ ,T_1^-]} 
     = 	N_1^{(r)}+N_2^{(s)}-N_2^{(r)}-N_3^{(s)}
		- (r,s\rightarrow \bar{r},\bar{s}), \nonumber\\
H_2 &\equiv& {[T_2^+ ,T_2^-]} 
	 = 	N_2^{(r)}+N_1^{(s)}-N_3^{(r)}-N_2^{(s)}
		- (r,s\rightarrow \bar{r},\bar{s}), \nonumber\\
H_3 &\equiv& {[T_3^+ ,T_3^-]} 
	 = 	N_2^{(r)}+N_3^{(s)}-N_3^{(\bar{r})}-N_4^{(s)}
		- (r,\bar{r},s\rightarrow \bar{r},r,\bar{s}). 
\end{eqnarray}
They are related to the Cartan generators in the orthonormal basis,
\begin{equation}
\vec{h}\equiv h_1\hat{e}_1+h_2\hat{e}_2+h_3\hat{e}_3
=[h_1,h_2,h_3],
\end{equation} 
as follows:
\begin{equation}
H_1=h_1-h_2,~H_2=h_2-h_3,~H_3=h_2+h_3. 
\end{equation}

For an ${\rm so}(6)$ irrep, its highest weight is 
\begin{equation}
\Lambda = (r_1^\dagger )^{a_1}(s_1^\dagger )^{a_2}(\bar{s}_{4}^\dagger )^{a_3}|0>, 
\end{equation}
where $a_{1,2,3}$ are non-negative integers. 
Action of the ${\rm so}(6)$ Cartan generators in the Dynkin basis on 
it yields
\begin{equation}
\vec{H}\Lambda=(a_1,a_2,a_3)\Lambda,
\end{equation}
and in the orthonormal basis 
\begin{equation}
\vec{h}\Lambda=[b_1,b_2,b_3]\Lambda,
\end{equation}  
 where
\begin{eqnarray}
b_1 &=& \frac 12(a_3+a_2)+a_1, \nonumber\\
b_2 &=& \frac 12(a_3+a_2), \nonumber\\
b_3 &=& \frac 12(a_3-a_2).
\end{eqnarray}

Inside the ${\rm so}(6)$ generators, the generators $T_{1,6}^\pm $ 
and $H_{1,6}$ form the ${\rm so}(4)$  Lie subalgebra and the generator 
$h_3$ is the generator of ${\rm so}(2)$ subalgebra. The other generators 
$T_{2,3,4,5}^\pm$ lie outside the subalgebra ${\rm so}(4)\times {\rm so}(2)$ 
and they are used to construct the Kostant operator of the quotient 
${\rm so}(6)/{\rm so}(4)\times {\rm so}(2)$.  

\subsection{Kernel solutions of the Kostant operator}

To construct the Kostant operator of the quotient 
${\rm so}(6)/{\rm so}(4)\times {\rm so}(2)$, 
the gamma matrices used here are
\begin{eqnarray}
%\label{002}
\gamma_2^{\pm} &=& \frac 12(\Gamma_1\pm i\Gamma_2),
\quad\quad
\gamma_3^{\pm} = \frac 12(\Gamma_3\pm i\Gamma_4),
\nonumber\\
\gamma_4^{\pm} &=& \frac 12(\Gamma_5\pm i\Gamma_6),
\quad\quad
\gamma_5^{\pm} = \frac 12(\Gamma_7\pm i\Gamma_8).
\end{eqnarray}
From the commutator of the generators of the quotient,
\begin{eqnarray}
\label{structure_constant2}
{[T_2^+ ,T_2^-]}&=& h_2-h_3,~~~~
{[T_3^+ ,T_3^-]} =  h_2+h_3,\nonumber\\
{[T_4^+ ,T_4^-]}&=& h_1-h_3,~~~~
{[T_5^+ ,T_5^-]} =  h_1+h_3,
\end{eqnarray}
the generators $T_{a=2,3,4,5}^\pm$ are not generated. 
The structure constants associated with these transformations are zero. 
Hence, the Kostant operator is just
\begin{equation}
\kostantop = \sum_{a=2}^5 (\gamma_a^+  T_a^- + \gamma_a^- T_a^+).
\end{equation}
A vector space of the Kostant operator is 
$\psi_\Lambda^{\pm}\equiv |\pm\pm\pm\pm\!\!>\!\otimes V_\Lambda$. 
Here, $V_{\Lambda}$ is the vector space of the ${\rm so}(6)$ irrep 
with the highest weight $\Lambda$. 
For the kernel solutions
\begin{equation}
\kostantop\psi_{\lambda_i}^{\pm} =0,
\end{equation}
where $\lambda_i$ is a weight in vector space $V_\Lambda$. 
Note that the derivation of the kernel solutions 
$\psi_{\lambda_3}^{+}$ and $\psi_{\lambda_8}^{+}$ in 
this quotient is not straightforward as the one in 
${\rm su}(5)/{\rm su}(4)\times {\rm u}(1)$. 
At a first glance, the following equations: 
\begin{eqnarray}
(T_2^+  + T_3^- - T_4^- + T_5^+ )
\psi_{\lambda_3}^{+}&=0, 
\nonumber\\
(T_2^-  - T_3^+ - T_4^+ - T_5^- )
\psi_{\lambda_8}^{+}&=0 
\end{eqnarray}
have kernel solutions
\begin{eqnarray}
\psi_{\lambda_3}^{+}&=|++-+>|0>,\nonumber\\
\psi_{\lambda_8}^{+}&=|+--->|0>.
\end{eqnarray} 
These solutions are true only when $a_1=a_2=a_3=0$. We fix this problem by 
twisting their spinor states and obtain the general kernel solutions in 
the positive spinor space:
\begin{eqnarray}
%\label{002}
\psi_{\lambda_1}^{+}&=&|++++> 
(r_1^\dagger )^{a_1}(s_1^\dagger )^{a_2}(\bar{s}_4^\dagger )^{a_3}|0>,
\nonumber\\
\psi_{\lambda_2}^{+}&=&|+++-> 
(r_1^\dagger )^{a_1}(s_2^\dagger )^{a_2}(\bar{s}_3^\dagger )^{a_3}|0>,
\nonumber\\
\psi_{\lambda_3}^{+}&=&|+---> 
(r_2^\dagger )^{a_1}(s_1^\dagger )^{a_2}(\bar{s}_4^\dagger )^{a_3}|0>, 
\nonumber\\
\psi_{\lambda_4}^{+}&=&|++--> 
(r_3^\dagger )^{a_1}(s_2^\dagger )^{a_2}(\bar{s}_4^\dagger )^{a_3}|0>,
\nonumber\\
\psi_{\lambda_5}^{+}&=&|+-++> 
(\bar{r}_1^\dagger )^{a_1}(s_3^\dagger )^{a_2}(\bar{s}_2^\dagger )^{a_3}|0>, 
\nonumber\\
\psi_{\lambda_6}^{+}&=&|+-+-> 
(\bar{r}_1^\dagger )^{a_1}(s_4^\dagger )^{a_2}(\bar{s}_1^\dagger )^{a_3}|0>,
\nonumber\\
\psi_{\lambda_7}^{+}&=&|+--+> 
(\bar{r}_3^\dagger )^{a_1}(s_1^\dagger )^{a_2}(\bar{s}_3^\dagger )^{a_3}|0>,
\nonumber\\
\psi_{\lambda_8}^{+}&=&|++-+> 
(\bar{r}_2^\dagger )^{a_1}(s_2^\dagger )^{a_2}(\bar{s}_3^\dagger )^{a_3}|0>, 
\end{eqnarray}
and in the negative spinor space:
\begin{eqnarray}
%\label{002}
\psi_{\lambda_1^\prime}^{-}&=&|-+++> 
(r_2^\dagger )^{a_1}(s_1^\dagger )^{a_2}(\bar{s}_2^\dagger )^{a_3}|0>, 
\nonumber\\
\psi_{\lambda_2^\prime}^{-}&=&|-++-> 
(\bar{r}_2^\dagger )^{a_1}(s_4^\dagger )^{a_2}(\bar{s}_3^\dagger )^{a_3}|0>,
\nonumber\\
\psi_{\lambda_3^\prime}^{-}&=&|-+-+> 
(\bar{r}_1^\dagger )^{a_1}(s_4^\dagger )^{a_2}(\bar{s}_2^\dagger )^{a_3}|0>, 
\nonumber\\
\psi_{\lambda_4^\prime}^{-}&=&|-+--> 
(\bar{r}_1^\dagger )^{a_1}(s_3^\dagger )^{a_2}(\bar{s}_1^\dagger )^{a_3}|0>,
\nonumber\\
\psi_{\lambda_5^\prime}^{-}&=&|--++> 
(r_2^\dagger )^{a_1}(s_3^\dagger )^{a_2}(\bar{s}_4^\dagger )^{a_3}|0>, 
\nonumber\\
\psi_{\lambda_6^\prime}^{-}&=&|--+-> 
(\bar{r}_2^\dagger )^{a_1}(s_2^\dagger )^{a_2}(\bar{s}_1^\dagger )^{a_3}|0>,
\nonumber\\
\psi_{\lambda_7^\prime}^{-}&=&|---+> 
(r_1^\dagger )^{a_1}(s_1^\dagger )^{a_2}(\bar{s}_3^\dagger )^{a_3}|0>,
\nonumber\\
\psi_{\lambda_8^\prime}^{-}&=&|----> 
(r_1^\dagger )^{a_1}(s_2^\dagger )^{a_2}(\bar{s}_4^\dagger )^{a_3}|0>.
\end{eqnarray}

To get the kernel solutions in terms of ${\rm so}(4)\times {\rm so}(2)$, 
it needs to act on them by the Cartan generators, which in the Dynkin basis are
\begin{eqnarray}
\label{diagonal_eq2}
D_1 &=& h_1-h_2 + {\frac 12\left(f_{+-1}^4[\gamma_4^+ ,\gamma_4^-] 
       + f_{+-1}^5[\gamma_5^+ ,\gamma_5^-] 
       - f_{+-2}^2[\gamma_2^+ ,\gamma_2^-] 
       - f_{+-2}^3[\gamma_3^+ ,\gamma_3^-]\right)}
\nonumber\\
	&=& H_1+ \frac 12(
	    \mathbbm{1}\otimes\sigma_3\otimes\sigma_3\otimes\mathbbm{1}
	   + \sigma_3\otimes\sigma_3\otimes\mathbbm{1}\otimes\mathbbm{1} 
	   - \mathbbm{1}\otimes\mathbbm{1}\otimes\mathbbm{1}\otimes\sigma_3
	   - \mathbbm{1}\otimes\mathbbm{1}\otimes\sigma_3\otimes\sigma_3),
\nonumber\\
 D_2 &=& h_1+h_2 + {\frac 12\left(f_{+-1}^4[\gamma_4^+ ,\gamma_4^-] 
       + f_{+-1}^5[\gamma_5^+ ,\gamma_5^-] 
       + f_{+-2}^2[\gamma_2^+ ,\gamma_2^-] 
       + f_{+-2}^3[\gamma_3^+ ,\gamma_3^-]\right)}
\nonumber\\
	&=& H_1+ H_2 + H_3 + \frac 12(
	    \mathbbm{1}\otimes\sigma_3\otimes\sigma_3\otimes\mathbbm{1}
	   + \sigma_3\otimes\sigma_3\otimes\mathbbm{1}\otimes\mathbbm{1} 
	   + \mathbbm{1}\otimes\mathbbm{1}\otimes\mathbbm{1}\otimes\sigma_3 \nonumber\\ 
	 &&~~~
	   + \mathbbm{1}\otimes\mathbbm{1}\otimes\sigma_3\otimes\sigma_3),
\nonumber\\
D_3 &=&  \frac 12 h_3 + {\frac 14\left(f_{+-3}^2[\gamma_2^+ ,\gamma_2^-] 
       + f_{+-3}^3[\gamma_3^+ ,\gamma_3^-] 
       + f_{+-3}^4[\gamma_4^+ ,\gamma_4^-]
       + f_{+-3}^5[\gamma_5^+ ,\gamma_5^-] \right)}
\nonumber\\
	&=& \frac 12 h_3 - \frac 14( 
	     \mathbbm{1}\otimes\mathbbm{1}\otimes\mathbbm{1}\otimes\sigma_3
	   - \mathbbm{1}\otimes\mathbbm{1}\otimes\sigma_3\otimes\sigma_3
	   + \mathbbm{1}\otimes\sigma_3\otimes\sigma_3\otimes\mathbbm{1}  
	   \nonumber\\ 
	 && ~~~
	 - \sigma_3\otimes\sigma_3\otimes\mathbbm{1}\otimes\mathbbm{1}).
\end{eqnarray}
The structure constants in (\ref{diagonal_eq2}) are read directly from 
(\ref{structure_constant2}). The generators $D_1$ and $D_2$ 
are the Cartan generators of ${\rm so}(4)$ and the generator 
$D_3$ is the Cartan generator of ${\rm so}(2)$. When the Cartan 
generators act on the kernel solutions, they yield
\begin{eqnarray}
\label{euler_multiplet_so4}
(D_1,D_2;D_3)\psi_{\lambda_1}^{+} 
 & = & (a_1,a_1+a_2+a_3+2;b_3/2)\psi_{\lambda_1}^{+},
\nonumber\\
(D_1,D_2;D_3)\psi_{\lambda_2}^{+}
 & = & (a_1+a_2+a_3+2,a_1;-b_3/2)\psi_{\lambda_2}^{+},
\nonumber\\
(D_1,D_2;D_3)\psi_{\lambda_3}^{+}
 & = & (-a_1,a_1+a_2+a_3;b_3/2)\psi_{\lambda_3}^{+},
\nonumber\\
(D_1,D_2;D_3)\psi_{\lambda_4}^{+}
 & = & (a_2,a_3;(b_1+2)/2)\psi_{\lambda_4}^{+},
\nonumber\\
(D_1,D_2;D_3)\psi_{\lambda_5}^{+}
 & = & (-a_1-a_2-a_3-2,-a_1;-b_3/2)\psi_{\lambda_5}^{+},
\nonumber\\
(D_1,D_2;D_3)\psi_{\lambda_6}^{+}
 & = & (-a_1,-a_1-a_2-a_3-2;b_3/2)\psi_{\lambda_6}^{+},
\nonumber\\
(D_1,D_2;D_3)\psi_{\lambda_7}^{+}
 & = & (a_3,a_2;-(b_1+2)/2)\psi_{\lambda_7}^{+},
\nonumber\\
(D_1,D_2;D_3)\psi_{\lambda_8}^{+}
 & = & (a_1+a_2+a_3,-a_1;-b_3/2)\psi_{\lambda_8}^{+},
\nonumber\\
(D_1,D_2;D_3)\psi_{\lambda_1^\prime}^{-}
 & = & (-a_1-a_3-1,a_1+a_2+1;-(b_2+1)/2)\psi_{\lambda_1^\prime}^{-},
\nonumber\\
(D_1,D_2;D_3)\psi_{\lambda_2^\prime}^{-}
 & = & (a_1+a_3+1,-a_1-a_2-1;-(b_2+1)/2)\psi_{\lambda_2^\prime}^{-},
\nonumber\\
(D_1,D_2;D_3)\psi_{\lambda_3^\prime}^{-}
 & = & (-a_1-a_3-1,-a_1-a_2-1;-(b_2+1)/2)\psi_{\lambda_3^\prime}^{-},
\nonumber\\
(D_1,D_2;D_3)\psi_{\lambda_4^\prime}^{-}
 & = & (-a_1-a_2-1,-a_1-a_3-1;(b_2+1)/2)\psi_{\lambda_4^\prime}^{-},
\nonumber\\
(D_1,D_2;D_3)\psi_{\lambda_5^\prime}^{-}
 & = & (-a_1-a_2-1,a_1+a_3+1;(b_2+1)/2)\psi_{\lambda_5^\prime}^{-},
\nonumber\\
(D_1,D_2;D_3)\psi_{\lambda_6^\prime}^{-}
 & = & (a_1+a_2+1,-a_1-a_3-1;(b_2+1)/2)\psi_{\lambda_6^\prime}^{-},
\nonumber\\
(D_1,D_2;D_3)\psi_{\lambda_7^\prime}^{-}
 & = & (a_1+a_3+1,a_1+a_2+1;-(b_2+1)/2)\psi_{\lambda_7^\prime}^{-},
\nonumber\\
(D_1,D_2;D_3)\psi_{\lambda_8^\prime}^{-}
 & = & (a_1+a_2+1,a_1+a_3+1;(b_2+1)/2)\psi_{\lambda_8^\prime}^{-}.
\end{eqnarray}
In case $a_1=a_2=a_3=0$, the kernel solutions (\ref{euler_multiplet_so4}) 
can be grouped in terms of ${\rm so}(4)$ dimensions as follows:
\begin{align}
({\bf 1},{\bf 1})_1 &\equiv \psi_{\lambda_4}^{+} \sim (~0,~0)_1,
\nonumber\\
({\bf 1},{\bf 3})_0 \equiv \left\{ 
\begin{array}{l}
\psi_{\lambda_1}^{+} \sim (~0,~2)_0 \\
\psi_{\lambda_8}^{+} \sim (~0,~0)_0\\
\psi_{\lambda_6}^{+} \sim (~0,-2)_0 
\end{array}
\right.,
\quad &\quad\quad
({\bf 3},{\bf 1})_0 \equiv 
\left\{
\begin{array}{l}
\psi_{\lambda_2}^{+} \sim (~2,~0)_0 \\
\psi_{\lambda_3}^{+} \sim (~0,~0)_0 \\
\psi_{\lambda_5}^{+} \sim (-2,~0)_0 
\end{array}
\right.,
\nonumber\\
({\bf 1},{\bf 1})_{-1}& \equiv \psi_{\lambda_7}^{+} \sim (~0,~0)_{-1}, 
\nonumber\\
({\bf 2},{\bf 2})_{1/2} \equiv 
\left\{
\begin{array}{l}
\psi_{\lambda_8^\prime}^{-} \sim (~1,~1)_{1/2} \\
\psi_{\lambda_6^\prime}^{-} \sim (~1,-1)_{1/2} \\
\psi_{\lambda_5^\prime}^{-} \sim (-1,~1)_{1/2} \\
\psi_{\lambda_4^\prime}^{-} \sim (-1,-1)_{1/2}  
\end{array}
\right.,
\quad &\quad
({\bf 2},{\bf 2})_{-1/2} \equiv 
\left\{
\begin{array}{l}
\psi_{\lambda_7^\prime}^{-} \sim (~1,~1)_{-1/2} \\
\psi_{\lambda_2^\prime}^{-} \sim (~1,-1)_{-1/2} \\
\psi_{\lambda_1^\prime}^{-} \sim (-1,~1)_{-1/2} \\
\psi_{\lambda_3^\prime}^{-} \sim (-1,-1)_{-1/2} 
\end{array}
\right..
\end{align}
Since the Dynkin labels $a_{1,2,3}$ are non-negative, the direct sum of 
the ${\rm so}(4)$ highest weights 
\begin{equation} 
\psi_{\lambda_4}^{+}
\oplus \psi_{\lambda_1}^{+}
\oplus \psi_{\lambda_2}^{+}
\oplus \psi_{\lambda_7}^{+}
\oplus \psi_{\lambda_8^\prime}^{-} 
\oplus \psi_{\lambda_7^\prime}^{-}
\end{equation}
forms the Euler number multiplet. In terms of its Dynkin labels, it reads:
\begin{eqnarray}
&(a_2,a_3)_{(b_1+2)/2}\oplus
(a_1,a_1+a_2+a_3+2)_{b_3/2}\oplus
(a_1+a_2+a_3+2,a_1)_{-b_3/2}\oplus
(a_3,a_2)_{-(b_1+2)/2}
\nonumber\\
&(a_1+a_2+1,a_1+a_3+1)_{(b_2+1)/2}\oplus
(a_1+a_3+1,a_1+a_2+1)_{-(b_2+1)/2}.
\end{eqnarray}

\section{Remarks}

Kernel solutions of the Kostant operator of the 8-dimensional 
quotients can be easily determined by the quantum mechanical method. 
The Euler number multiplet obtained in terms of the diagonal 
subalgebra is the direct sum of the highest weights of the kernel solutions, 
which appear only once. The Euler number multiplets presented in this paper 
are exactly the same as derived by using the Weyl group elements 
of ${\rm su}(5)$ and ${\rm so}(6)$ that are not in their subalgebras \cite{GKRS}. 
The lowest line of the Euler number multiplet for the quotient 
${\rm su}(5)/{\rm su}(4)\times {\rm u}(1)$ is
\begin{equation}
1_1\oplus 4_{1/2} \oplus 6_0 \oplus 4_{-1/2} \oplus 1_{-1},
\end{equation}  
and for the quotient ${\rm so}(6)/{\rm so}(4)\times {\rm so}(2)$ 
\begin{equation}
(1,1)_1\oplus (2,2)_{1/2} \oplus (3,1)_0 \oplus (1,3)_0 
\oplus (2,2)_{-1/2} \oplus (1,1)_{-1}.
\end{equation} 
There are several ways to interpret these Euler number 
multiplets. If ${\rm so}(2)$, which is locally isomorphic to 
${\rm u}(1)$, is viewed as a light-cone little group of 
${\rm ISO}(3,1)$, then they correspond to degrees of freedom 
of $N=4$ Yang-Mills massless representation in 3+1 space-time. 
Similarly, if ${\rm so}(6)$, which is locally isomorphic 
to ${\rm su}(4)$, is viewed as a light-cone little group of 
${\rm ISO}(7,1)$, then they correspond to degrees of freedom 
of the massless representation in 7+1 space-time. Lastly, 
if ${\rm so}(6)\times {\rm so}(2)$, which is locally isomorphic to 
${\rm su}(4)\times {\rm u}(1)$, is viewed as a subgroup of 
${\rm SO}(6,2)$, the anti-de Sitter group and the conformal group, 
then they correspond to the massless representations in the 
6+1 and 5+1 space-time, respectively.            

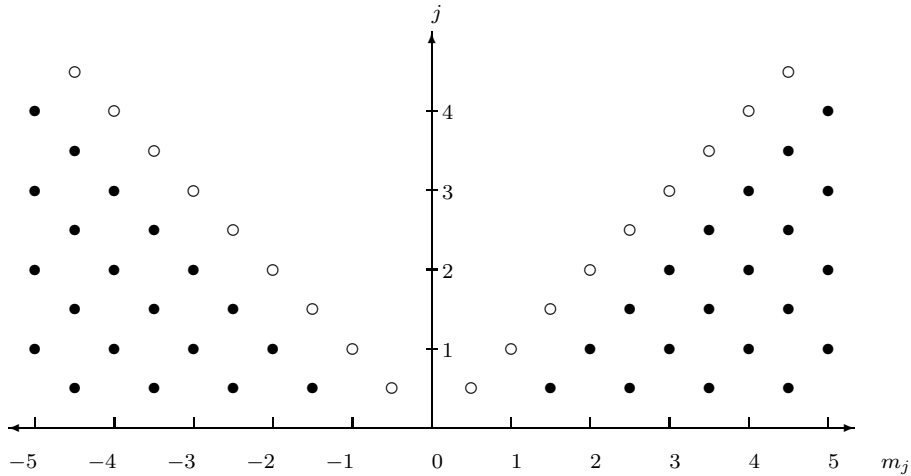
\begin{figure}[h]
\begin{center}
\begin{picture}(360,180)(0,0)

\footnotesize

\put(20,0){$-5$}
\put(50,0){$-4$}\put(80,0){$-3$}
\put(110,0){$-2$}\put(140,0){$-1$}
\put(180,0){0}\put(210,0){1}
\put(240,0){2}\put(270,0){3}
\put(300,0){4}\put(330,0){5}
\put(350,0){$m_j$}

\put(180,15){\vector(1,0){160}}
\put(180,15){\vector(-1,0){160}}
\multiput(30,15)(30,0){11}{\line(0,1){4}}
\put(180,15){\vector(0,1){150}}
\put(180,170){$j$}
\multiput(178,45)(0,30){4}{\line(1,0){4}}
\put(184,42){1}
\put(184,72){2}
\put(184,102){3}
\put(184,132){4}

\multiput(45,30)(30,0){4}{\circle*{4}}
\put(165,30){\circle{4}}
\put(195,30){\circle{4}}
\multiput(225,30)(30,0){4}{\circle*{4}}

\multiput(30,45)(30,0){4}{\circle*{4}}
\put(150,45){\circle{4}}
\put(210,45){\circle{4}}
\multiput(240,45)(30,0){4}{\circle*{4}}

\multiput(45,60)(30,0){3}{\circle*{4}}
\put(135,60){\circle{4}}
\put(225,60){\circle{4}}
\multiput(255,60)(30,0){3}{\circle*{4}}

\multiput(30,75)(30,0){3}{\circle*{4}}
\put(120,75){\circle{4}}
\put(240,75){\circle{4}}
\multiput(270,75)(30,0){3}{\circle*{4}}

\multiput(45,90)(30,0){2}{\circle*{4}}
\put(105,90){\circle{4}}
\put(255,90){\circle{4}}
\multiput(285,90)(30,0){2}{\circle*{4}}

\multiput(30,105)(30,0){2}{\circle*{4}}
\put(90,105){\circle{4}}
\put(270,105){\circle{4}}
\multiput(300,105)(30,0){2}{\circle*{4}}

\put(45,120){\circle*{4}}
\put(75,120){\circle{4}}
\put(285,120){\circle{4}}
\put(315,120){\circle*{4}}

\put(30,135){\circle*{4}}
\put(60,135){\circle{4}}
\put(300,135){\circle{4}}
\put(330,135){\circle*{4}}

\put(45,150){\circle{4}}
\put(315,150){\circle{4}}

\end{picture}

\end{center}

\caption{The ${\rm so}(2,1)$ weight diagram associated with 
the discrete representations. Open and solid circles along 
a horizontal line are the ${\rm so}(2,1)$ weights of a $j$ 
representation. In each horizontal line, only the open circles, 
the lowest weight in the $V_j^+$ and the highest weight in 
the $V_j^-$, are the non-trivial kernel solutions of the 
Kostant operator of the quotient ${\rm so}(2,1)/{\rm so}(2)$.}
\label{Fig3}
\end{figure}

The Kostant operator can be extended from a compact Lie algebra 
to a non-compact one. Methods to construct the Kostant operator 
are similar in both the compact and the non-compact Lie algebras. The 
simplest quotient of the non-compact Lie algebras is 
${\rm so}(2,1)/{\rm so}(2)$. For details of the ${\rm so}(2,1)$ 
generators, commutation relations and representations, 
see \cite{Wybourne}. The Kostant operator,
\begin{equation}
\kostantop= \sigma^+ T^- + \sigma^- T^+,
\end{equation}
acts on its vector space $\psi_j^\pm=|\pm>|j,m_j>$, 
where in each discrete representation $j$, $|m_j|\geq j$. 
Its non-trivial kernel solutions, whose corresponding 
states are shown as open circles in 
figure \ref{Fig3}, are
\begin{equation}
\psi_j^+ = |+>|j,-j>,
\quad\quad
\psi_j^- = |->|j,j>.
\end{equation}
These solutions are similar to the kernel solutions of 
${\rm su}(2)/{\rm u}(1)$. Another interesting non-compact 
Lie algebra is ${\rm so}(4,2)$, the conformal group in 
the 3+1 space-time,  whose spinors are twistors \cite{Penrose}. 
For the case ${\rm so}(4,2)/{\rm so}(4)\times {\rm so}(2)$, 
it is found that the lowest line of its Euler number 
multiplet for the discrete representation is similar to that 
of ${\rm so}(6)/{\rm so}(4)\times {\rm so}(2)$.   

Finally, it is hoped that the constructions of 
the Kostant operators and the derivations of their 
kernel solutions presented here will be 
useful when someone wants to oxidize a low-dimensional 
field theory to a higher-dimensional one or to reduce 
a high-dimensional field theory to a lower-dimensional 
one \cite{ABR}, or even to connect the Kostant operators 
to the string theory \cite{Agricola,Metsaev}.

\end{document}